\documentclass{article}

\usepackage[preprint, nonatbib]{neurips_2023}

\usepackage[utf8]{inputenc} 
\usepackage[T1]{fontenc}    
\usepackage{hyperref}       
\usepackage{url}            
\usepackage{booktabs}       
\usepackage{amsfonts}       
\usepackage{nicefrac}       
\usepackage{microtype}      
\usepackage{xcolor}         

\title{AI Wellbeing}

\author{%
  Simon Goldstein\thanks{Equal contribution.}\\
  Australian Catholic University\\
  Center for AI Safety\\
  \And
  Cameron Domenico Kirk-Giannini$^{*}$ \\
  Rutgers University--Newark\\
  Center for AI Safety\\
}

\begin{document}

\maketitle

\begin{abstract}
Under what conditions would an artificially intelligent system have wellbeing? Despite its obvious bearing on the ethics of human interactions with artificial systems, this question has received little attention. Because all major theories of wellbeing hold that an individual’s welfare level is partially determined by their mental life, we begin by considering whether artificial systems have mental states. We show that a wide range of theories of mental states, when combined with leading theories of wellbeing, predict that certain existing artificial systems have wellbeing. While we do not claim to demonstrate conclusively that AI systems have wellbeing, we argue that our metaphysical and moral uncertainty about AI wellbeing requires us dramatically to reassess our relationship with the intelligent systems we create.
\end{abstract}

We recognize one another as beings for whom things can go well or badly,
beings whose lives may be better or worse according to the balance they
strike between goods and ills, pleasures and pains, desires satisfied
and frustrated. In our more broad-minded moments, we are willing to
extend the concept of wellbeing also to nonhuman animals, treating them
as independent bearers of value whose interests we must consider in
moral deliberation.\addtocounter{footnote}{-1}\footnote{Following Heathwood (2008) and others, we
  understand \emph{wellbeing} to be a kind of non-instrumental goodness
  for: what contributes to an entity's wellbeing is what is
  non-instrumentally good for it. Wellbeing is morally significant in
  the sense that entities that have wellbeing have a distinctive moral
  status which obliges us during moral deliberation to consider which
  outcomes are good or bad for them. While there is a sense in which
  growing is non-instrumentally good for plants, for example, we do not
  think this entails that they have wellbeing.} But most people, and
perhaps even most philosophers, would reject the idea that fully
artificial systems, designed by human engineers and realized on computer
hardware, may similarly demand our moral consideration. Even many who
accept the possibility that humanoid androids in the distant future will
have wellbeing would resist the idea that the same could be true of
existing AI systems today.

Perhaps because the creation of artificial systems with wellbeing is
assumed to be so far off, little philosophical attention has been
devoted to the question of what such systems would have to be like. In
what follows, we suggest a surprising answer to this question: when one
integrates leading theories of mental states like belief, desire, and
pleasure with leading theories of wellbeing, one is confronted with the
possibility that the technology already exists to create AI systems with
wellbeing. We argue that a new type of AI system -- the \emph{artificial
language agent} -- has wellbeing. Artificial language agents augment
large language models (LLMs) with the capacity to observe, remember, and
form plans. We also argue that the possession of wellbeing by artificial
language agents does not depend on them being phenomenally conscious.
Given that artificial language agents demonstrate an improved capacity
for long-term planning compared to other contemporary AI systems, we
expect that they will become increasingly common in the near future. Far
from a topic for speculative fiction or future generations of
philosophers, then, AI wellbeing is a pressing issue.

We begin by introducing the architecture of artificial language agents
and the machine learning models on which they are based (Section 1). We
then consider whether artificial language agents have beliefs and
desires (Section 2), and whether they can experience pleasure (Section
3). The answers to these questions inform our subsequent discussion of
whether artificial language agents have wellbeing according to hedonism
(Section 3), desire-satisfactionism (Section 4), and objective list
theories (Section 5). Our thesis is potentially threatened by the idea
that phenomenal consciousness is necessary for being a welfare subject,
so we carefully explore the plausibility of this idea (Section
6).\footnote{We use the terms \emph{wellbeing} and \emph{welfare} as
  synonyms. A \emph{welfare subject} is an entity that possesses welfare
  or wellbeing. A being's \emph{welfare level} is the amount of welfare
  or wellbeing it possesses. A \emph{welfare good} is something which
  contributes to the welfare level of the welfare subjects that possess
  it.} We conclude by replying to some potential objections (Section 7)
and discussing the implications of our uncertainty about whether systems
like artificial language agents have wellbeing (Section 8).

\section{Artificial Language Agents}

Artificial language agents (from now on simply \emph{language agents})
are our central focus in what follows because this will afford us the
strongest case that existing AI systems have wellbeing. Language agents
are built by wrapping an LLM in a larger functional architecture that
allows the system to engage in long term planning. We'll start by
briefly explaining how LLMs work, and then turn to language agents in
detail.

At the cognitive core of every language agent is a large language model.
An LLM is an artificial neural network designed to generate coherent
text responses to text inputs. Large language models exploded into
public attention in 2022 with the launch of OpenAI's ChatGPT. Systems
like GPT-3.5, the model underlying ChatGPT, fluently respond to a wide
range of text prompts. They can answer factual questions, write prose in
any genre, and generate working code in many programming
languages.\footnote{It is beyond the scope of our discussion to describe
  the technical details underwriting the capabilities of LLMs. But it is
  worth mentioning that they depend on an architectural innovation
  called the \emph{transformer}, which improves neural network models'
  ability to keep track of complex dependency relationships between
  their inputs (for details, see Vaswani et al. 2017).}

Think of the LLM at the center of a language agent as its cerebral
cortex: it performs most of the agent's cognitive processing tasks. In
addition to the LLM, however, a language agent has files that record its
beliefs, desires, plans, and observations in natural language. The
programmed architecture of a language agent gives these beliefs,
desires, plans, and observations their functional roles by specifying
how they are processed by the LLM in determining how the agent acts. The
agent observes its environment, summarizes its observations using the
LLM, and records the summary in its beliefs. Then it calls on the LLM to
form a plan of action based on its beliefs and desires. In this way, the
cognitive architecture of language agents is familiar from folk
psychology.

For concreteness, consider the language agents developed by Park et
al. (2023). These agents live in a simulated world called `Smallville',
with which they can observe and interact via natural-language
descriptions of what they see and how they choose to act. Each agent is
given a text backstory that defines their occupation, relationships,
and goals. As they navigate the world of Smallville, their experiences
are added to a ``memory stream.'' The program that defines each agent
feeds important memories from each day into the underlying language
model, which generates a plan for the next day. Plans determine how an
agent acts but can be revised on the fly on the basis of events that
occur during the day.

More carefully, the language agents in Smallville choose how to behave
by \emph{observing, reflecting,} and \emph{planning}. As each agent
navigates the world, all of its observations are recorded in its memory
stream in the form of natural language statements about what is going on
in its immediate environment. Because the agent's memory stream is long,
agents use the LLM (in this case, gpt3.5-turbo) to assign importance
scores to their memories and to determine which memories are relevant to
their situation. Then the agents reflect: they query the LLM to make
important generalizations about their values, relationships, and other
higher-level representations. Finally, they plan: each day, agents use
the LLM to form and revise a detailed plan of action based on their
memories of the previous day together with their other relevant and
important beliefs and desires. In this way, the LLM engages in practical
reasoning, developing plans that promote the agent's goals given the
agent's beliefs. Plans are entered into the memory stream alongside
observations and reflections and shape the agent's behavior throughout
the day.

Large language models are good at reasoning and producing fluent text.
By themselves, however, they can't form memories or execute long-term
plans. Language agents build on the reasoning abilities of LLMs to
create full-fledged planning agents.

Besides the agents developed by Park et al., other potential examples of
language agents include AutoGPT\footnote{Project available at
  \textless https://github.com/Significant-Gravitas/Auto-GPT\textgreater.},
BabyAGI\footnote{Project available at
  \textless https://github.com/yoheinakajima/babyagi\textgreater.},
Voyager\footnote{See Wang et al. (2023).}, and SPRING\footnote{See Wu et
  al. (2023).}. Each of these systems has a distinct architecture, and
the differences between them may at times be relevant to our discussion
in what follows. Unless we explicitly flag differences, the term
``language agents'' should be understood to denote agents with
architectures very similar to the one described in Park et al.

Note that, while existing language agents are reliant on text-based
observation and action spaces, the technology already exists to
implement language agents in real-world settings. The rise of multimodal
language models like GPT-4, which can interpret image as well as text
inputs, and the possibility of using such language models to control a
mobile robotic system, as in Google's PaLM-E (Dreiss et al. 2023), mean
that the possible applications of language agents are extremely diverse.

\section{Belief and Desire}

Can language agents have beliefs and desires? To answer this question,
we consider a range of theories of belief and desire which place
increasingly strong demands on the internal structure of the believing
agent, starting with dispositionalism and interpretationism and ending
with representationalism. As we will see, almost all of the theories we
canvass suggest that language agents and related systems can have
beliefs and desires.

According to the dispositionalist, to believe or desire that \emph{P} is
to possess a suitable suite of dispositions across a variety of actual
and possible circumstances. The dispositions constitutive of a mental
state may, depending on the particular dispositionalist account, include
dispositions to behave, dispositions to token other mental states
(\emph{cognitive} dispositions), and dispositions to have phenomenally
conscious experiences (\emph{phenomenal} dispositions).\footnote{The
  view that mental states like belief and desire are constituted
  exclusively by behavioral dispositions is a form of
  \emph{behaviorism}. We do not focus on behaviorism in what follows
  because it is not a popular position among philosophers or cognitive
  scientists. Note, however, that behaviorism entails that artificial
  systems can have beliefs and desires. For more on behaviorism, see
  e.g. Ryle (1949) and Place (1956, 2002).} We will refer to
dispositionalist accounts which do not appeal to phenomenal dispositions
as versions of \emph{narrow dispositionalism} and dispositionalist
accounts which do appeal to phenomenal dispositions as versions of
\emph{wide dispositionalism}. Narrow dispositionalism about belief
and/or desire has influentially been defended by Stalnaker (1984) and
Marcus (1990). Indeed, narrow dispositionalism is so popular that
Schroeder (2004), one of its opponents, refers to it as the `standard
theory' of desire. As Stalnaker formulates the view:

\begin{quote}
``To desire that \emph{P} is to be disposed to act in ways that would
tend to bring it about that \emph{P} in a world in which one's beliefs,
whatever they are, were true. To believe that \emph{P} is to be disposed
to act in ways that would tend to satisfy one's desires, whatever they
are, in a world in which \emph{P} (together with one's other beliefs)
were true.'' (1984, 15)
\end{quote}

And Marcus writes:

\begin{quote}
``...\emph{x} believes that \emph{S} just in case under certain
\emph{agent-centered} circumstances including \emph{x}\textquotesingle s
desires and needs as well as \emph{external circumstances}, \emph{x} is
disposed to act as if \emph{S}, that actual or non-actual state of
affairs, obtains.'' (1990, 140; emphasis in original)
\end{quote}

As these quotes suggest, many dispositionalists hold that the
dispositional profile of belief cannot be specified without reference to
the dispositional profile of desire, and vice versa. So, to determine
whether language agents have beliefs and desires, the dispositionalist
must check whether their total set of behavioral dispositions is that of
a being which acts on its beliefs to satisfy its desires.

In the case of a language agent, the best candidate for the state of
believing that \emph{P} is the state of having a declarative sentence
with \emph{P} as its content written in its memory stream. This state is
accompanied by the right kinds of verbal and nonverbal behavioral
dispositions to count as a belief that \emph{P}, and, given the
functional architecture of the system, also the right kinds of cognitive
dispositions. The best candidate for the state of desiring \emph{P} is
having a declarative sentence with \emph{You desire that P} as its
content in the memory stream. Such sentences can be found in each
agent's initial description. For example, one of Park et al.'s language
agents had an initial description that included the goal of planning a
Valentine's Day party. This goal was entered into the agent's planning
module along with a summary of important events from the memory stream.
The result was a complex pattern of behavior. The agent met with every
resident of Smallville, inviting them to the party and asking them what
kinds of activities they would like to include. Their feedback was
incorporated into the party planning. This kind of complex behavior is
part of a disposition to act in ways that would tend to bring about a
successful Valentine's Day party, given the agent's observations about
the world they inhabit.

Desire may also involve other cognitive dispositions. For example,
philosophers like Scanlon (1998) and Sinhababu (2017) have suggested
that one role of desire is to influence attention. Artificial systems
could also have states which influence their attention in the relevant
way. Indeed, Park et al.'s (2023) language agents use a special process
of considering their goals when deciding whether to direct attention
toward a novel observation in the memory stream.

The fact that narrow dispositionalists tend to reduce belief and desire
simultaneously to behavioral dispositions brings their view close to
another tradition in philosophy of mind: interpretationism.\footnote{More
  carefully, narrow dispositionalists reduce belief and desire to a
  combination of behavioral and cognitive dispositions, but the mental
  states in terms of which these cognitive dispositions are specified
  are, at least for Stalnaker, themselves picked out in terms of the
  roles they play in explaining behavior.} Interpretationists like
Donald Davidson and Daniel Dennett hold that what it is to have beliefs
and desires is for one's behavior (both verbal and nonverbal) to be
suitably interpretable as rational given those beliefs and desires. Thus
Davidson remarks that ``In interpreting utterances from scratch---in
\emph{radical} interpretation---we must somehow deliver simultaneously a
theory of belief and a theory of meaning'' (1974, 312), that ``The only
basis for a theory of meaning is the whole fabric of belief as evinced
in a system of behavior'' (1970/2020, 113), and that ``What a fully
informed interpreter could learn about what a speaker means is all there
is to learn; the same goes for what the speaker believes'' (1986, 315).
Similarly, Dennett holds that ``any system\ldots{} whose behavior is
well predicted by {[}treating it as a rational agent with beliefs and
desires{]} is in the fullest sense of the word a believer'' (1981, 15).

Interpretationism differs from dispositionalism in its emphasis on
interpretation. It is similar to some versions of dispositionalism,
however, in holding that the relevant conditions for belief and desire
are publicly observable. They must be the sorts of conditions an
interpreter could notice without knowing the internal cognitive
structure of the believer. Again, the agent who plans a Valentine's Day
party is ripe for interpretationist analysis. Their behavior would be
very hard to explain without referencing the goal of a Valentine's Day
party. Short of that goal, there is little to unify the large list of
conversations the agent has throughout their day, and little to explain
why the party gradually took shape.\footnote{Child (1994, 47)
  distinguishes between \emph{constitutive} and \emph{non-constitutive}
  versions of interpretationism, where the constitutive
  interpretationist holds that being suitably interpretable as having
  beliefs constitutes having those beliefs and the non-constitutive
  interpretationist holds only that a thing is suitably interpretable as
  having beliefs just in case it has those beliefs. Since our question
  concerns whether artificial systems can have beliefs, we focus on the
  weaker, non-constitutive thesis.

  Child also distinguishes between \emph{pure} interpretationism and
  \emph{supplemented} interpretationism. Pure interpretationism is the
  view that being interpretable as having certain beliefs is itself
  sufficient for having those beliefs, while supplemented
  interpretationism is the view that being interpretable as having
  certain beliefs is only sufficient for having those beliefs when
  certain further background conditions obtain. Various difficult cases
  could motivate supplemented interpretationism. For example, some
  versions of pure interpretationism might predict that thermometers
  have beliefs and desires. In response, interpretationists like Dennett
  hold that being interpretable as having beliefs and desires is only
  sufficient for possessing beliefs and desires when attributing folk
  psychological concepts to a system allows us to explain it
  \emph{better} than by thinking of it as a physical system or an
  artifact.}

Apart from their differences in emphasis, interpretationism and narrow
dispositionalism agree that what counts in attributing beliefs and
desires to an agent is how they are disposed to act across a variety of
possible circumstances, where \emph{act} is understood to include verbal
behavior. Both views are representationally lightweight: no particular
cognitive or biological internal structure is necessary. For this
reason, narrow dispositionalism and interpretationism both predict that
a range of artificial systems could have beliefs and desires. On the
theories of Stalnaker or Marcus, for example, a simple reinforcement
learning agent could be said to have beliefs if it responded
differentially to changes in its environment in a way that promoted
achieving its goals. And for the interpretationist, we could apply the
methods of radical interpretation to the linguistic and nonlinguistic
behavior of a language agent to determine what it believes. We conclude
that all viable narrow dispositionalist and interpretationist theories
of belief and desire predict that language agents have beliefs and
desires, and many also predict that simpler systems which do not produce
natural-language outputs have beliefs and desires.\footnote{See Butlin
  (2023) for further discussion of whether AIs trained in reinforcement
  learning count as genuine agents.}

We turn now to wide dispositionalism, the view that to believe or desire
that \emph{P} is to possess a suite of dispositions including phenomenal
dispositions. Wide dispositionalism has recently been championed by
Schwitzgebel (2002), who argues that belief is individuated in terms of
all three types of dispositions: behavior, cognitive, and
phenomenal.\footnote{Note that, while Schwitzgebel holds that belief is
  partially individuated in terms of phenomenal dispositions, to our
  knowledge he offers no argument that this view is explanatorily
  superior to narrow dispositionalist views which bring in cognitive
  dispositions. Both sorts of dispositionalist views have the resources
  to respond to various objections indicating that mental states cannot
  be understood exclusively in terms of behavioral dispositions.
  Accordingly, we wonder whether appealing to phenomenal dispositions in
  explicating belief and desire is well motivated.} For Schwitzgebel,
beings which share some but not all of the dispositional profile
associated with paradigm cases of belief are borderline cases of
believers. On Schwitzgebel's view, then, in order for artificial systems
to determinately be believers, they would need to have phenomenal
experiences. Even if artificial systems cannot be phenomenally
conscious, however, Schwitzgebel's view predicts that they can be
borderline cases of believers if they have the right behavioral and
cognitive dispositions.

Along similar lines, we have the \emph{hedonic theory} of desire, which
is a version of wide dispositionalism according to which an agent
desires \emph{P} just in case it is disposed to experience pleasure from
it seeming that \emph{P} (Mill 1863; Strawson 1994; Schroeder 2004, 38).
If the hedonic theory of desire is correct and artificial systems cannot
be phenomenally conscious, then it would seem that they cannot have
desires.

While wide dispositionalism is a coherent position, most theories of
belief and desire suggest that there is no necessary connection between
belief or desire and phenomenal consciousness. And arguably this is as
it should be. We think it is conceivable that an agent could have
Kantian moral desires --- desires that motivated it to act `out of duty'
without pleasure. When it comes to the relationship between desire and
phenomenal consciousness, Schroeder (2004, 26) points out that:

\begin{quote}
``The standard theory of desire holds that desires do not depend upon
consciousness for their existence but upon motivational structures, and
the involvement of a desire in consciousness is not necessary for it to
carry out its functional role. Accordingly, a desire is a desire whether
it is part of consciousness or not.''
\end{quote}

A similar point could be made about belief. With few exceptions,
theories of the nature of belief have held that a belief need not be
accompanied by any phenomenal state. Again, this is arguably as it
should be. If an advanced species of aliens made contact with humanity,
we would plausibly be able to know that members of this species had
beliefs and desires even if we were uncertain about whether their
cognitive apparatus had a structure appropriate to generate phenomenal
consciousness.

Though wide dispositionalism ties belief to states beyond observable
behavior, it still places few substantive requirements on the causal or
functional organization of a believing or desiring agent's cognitive
apparatus. We turn now to representationalism, a view which holds that
belief and desire are constituted by factors more clearly ``inside the
head'' of the believing agent. Representationalism deserves special
emphasis in the context of our discussion because ``probably the
majority of contemporary philosophers of mind adhere to some form of
representationalism about belief'' (Schwitzgebel 2011, 15).

Representationalists hold that to believe or desire that \emph{P} is to
token a representational vehicle with the appropriate causal powers
having \emph{P} as its content. For example, Fodor (1987, 10) proposes
that a psychological theory posits beliefs and desires just in case ``it
postulates states \ldots{} satisfying the following conditions:

\begin{quote}
(i) They are semantically evaluable.

(ii) They have causal powers.

(iii) The implicit generalizations of commonsense belief/desire
psychology are largely true of them.''\footnote{For further discussion
  of representationalism about desire (for example, the thesis that one
  desires \emph{P} just in case one has a mental representation with the
  content that \emph{P} that motivates one to bring about \emph{P}), see
  Block (1986), Cummins (1989), Harman (1973), Millikan (1984), and
  Papineau (1987).}
\end{quote}

It is hard to resist the conclusion that language agents have beliefs
and desires in the Fodorian sense. Park et al.'s (2023) agents, for
example, have memories which consist of text files containing natural
language sentences specifying what they have observed and what they
want. Natural language sentences are clearly semantically evaluable, and
the fact that a given sentence is in a given agent's memory plays a
direct causal role in shaping its behavior. It is possible to reason
about the behavior of a language agent on the basis of its beliefs and
desires. Language agents satisfy the language of thought hypothesis:
their language of thought is English!

We haven't yet mentioned functionalism, the view that mental states like
belief and desire are individuated by the roles they play in larger
systems. This is because without further specification of the relevant
functional role, functionalism does not answer the question of whether
language agents can have beliefs and desires. For example, narrow
dispositionalism is a functionalist theory which specifies the relevant
functional role entirely in terms of behavioral and cognitive
dispositions, and it predicts that systems like language agents can have
beliefs and desires. On the other hand, psychofunctionalism specifies
the functional roles that individuate belief and desire in terms of an
empirical theory of human cognition. Given how much the internal
constitution of artificial systems like language agents differs from
that of humans, the psychofunctionalist may not grant that language
agents have beliefs and desires.\footnote{This is an instance of the
  observation, which some have taken to constitute a serious objection,
  that psychofunctionalism is ``chauvinistic'' --- see Block (1978) for
  discussion.}

In the present context, two functionalist proposals are particularly
worth discussing. The first, due to Putnam (1960, 1967), identifies a
creature's mental states with states of the Turing machine describing
that creature's cognitive apparatus. This sort of functionalist picture
closely approximates narrow dispositionalism in so far as it holds that
the state of believing or desiring that \emph{P} is individuated by its
relationship with other cognitive states and potentially also sensory
inputs and behavioral outputs. It therefore leaves open the possibility
that artificial systems like language agents can have beliefs and
desires.

The second functionalist proposal, associated with Lewis (1972), seeks
to identify mental states like belief and desire by first constructing a
set of Ramsey sentences from the platitudes of folk psychology and then
finding the states, whatever they are, that witness the Ramsey
sentences.\footnote{A Ramsey sentence is a quantified sentence
  describing the theoretical role of a mental state without reference to
  mentalistic language.} Since the natural-language representations in
systems like language agents are designed to function in accordance with
the platitudes of folk psychology, this sort of functionalism would seem
to predict more or less directly that language agents and similar
systems can have beliefs and desires.

We conclude that a wide range of accounts of the nature of belief and
desire entail that systems like language agents can have beliefs and
desires.

\section{Pleasure and Hedonism}

We turn now from belief and desire to pleasure, and in this context also
from focusing on issues in the philosophy of mind to focusing on issues
is the philosophy of wellbeing. There are three main theories of
wellbeing: hedonism, desire satisfactionism, and objective list
theories. According to hedonism, wellbeing is a function of pleasure and
pain. Your life goes well to the extent that you have many pleasurable
experiences and few painful ones. According to desire satisfactionism,
wellbeing is a function of your desires. Your life goes well to the
extent that many of your desires are satisfied. According to objective
list theories, wellbeing is determined by the possession of objectively
valuable things. A life is good, on this view, to the extent that it is
filled with knowledge, friendship, achievement, and other kinds of human
flourishing.

We will consider each theory in turn, beginning in this section with
hedonism. At first, it might seem that hedonism rules out the
possibility that artificial systems could have wellbeing because it
reduces this question to the question of whether they can experience
pleasure. We argue that this conclusion is too hasty: it is possible
that language agents have wellbeing even if hedonism is true. In the
following two sections, we argue that desire satisfactionism and
objective theories of wellbeing make it even likelier that language
agents have wellbeing.

Hedonism says that pleasure is what makes one's life go well, and pain
is what makes one's life go badly. To determine whether language agents
have wellbeing, on this view, we must determine whether they feel
pleasure and pain. This in turn depends on the nature of pleasure and
pain.

Before getting into details, it is worth clarifying two things. First,
most hedonists include a wide range of negative experiences under the
heading of pain. Nausea, depression, and itching might not be called
`painful' in ordinary speech, but they are unpleasant experiences (or
`displeasures'), and hedonists count them as making life worse (Gregory
2015, 115). Second, it is worth distinguishing between two notions of
pleasure. \emph{Sensory pleasure} refers to pleasurable experiences.
\emph{Propositional pleasure} refers to taking pleasure in states of
affairs, like when we are pleased that the store is open. In principle,
one could be a hedonist about wellbeing in the sense of sensory
pleasure, propositional pleasure, or both. But we focus on sensory
pleasure, since it is the harder case for the thesis that artificial
systems can have wellbeing.

There are two main theories of sensory pleasure and displeasure.
According to \emph{phenomenal theories} (see Bramble 2013, Kagan 1992),
sensory pleasures are phenomenal states. According to \emph{attitudinal
theories}, sensory pleasure is explained in terms of propositional
pleasure: what it is for a sensation to be pleasant is for its subject
to take propositional pleasure in having that sensation (see Alston
1967, Brandt 1966, Schroeder 2004, Feldman 2004, and Heathwood 2016). If
phenomenal theories are correct, then language agents probably do not
have hedonistic wellbeing. But if attitudinal theories are correct,
language agents may have wellbeing.

One phenomenal theory of sensory pleasure is the \emph{distinctive
feeling theory}. The distinctive feeling theory says that there is a
particular phenomenal experience of pleasure that is common to all
pleasant activities. For example, Bramble (2013, 210) argues that the
felt quality of a pleasurable experience is a particular sensation that
permeates the experience, can come in very low intensities, and is
scattered in finely discriminable bits throughout the experiential
field, in a way that can be elusive. We see little reason why language
agents would have representations with this kind of structure. So if
this theory of pleasure were correct, then hedonism would predict that
language agents do not have wellbeing.

In contrast to the distinctive feeling theory, the \emph{hedonic tone}
theory says that various pleasurable experiences share a common aspect.
For example, Kagan (1992) suggests that sensory experiences can vary
along a dimension of how pleasurable they are, analogously to how
auditory experiences vary in how loud they are. Just as there is no
specific felt quality of volume, so there is no specific felt quality of
pleasure. Nonetheless, volume and pleasure describe important aspects of
experience. Again, we suspect that language agent representations lack
this structure, and so if sensory pleasure involves hedonic tone, then
language agents do not have sensory pleasure or pain.

The main alternative to phenomenal theories of sensory pleasure is
attitudinal theories. In fact, Bramble (2016) notes that most
philosophers of wellbeing favor attitudinal over phenomenal theories of
sensory pleasure. This position is motivated by the apparent
heterogeneity of pleasure: a wide range of disparate experiences are
pleasant, including the warm relaxation of soaking in a hot tub, the
taste of chocolate cake, the excitement of winning an award, and the
challenge of completing a crossword.

Desire-based theories of pleasure say that experiences are pleasant when
they are desired. There are a variety of theories in this tradition:
Alston (1967, 365) holds that the experience must be desired for its
``felt quality''; Brandt (1966, 38) holds that one must desire that the
experience continue; and Heathwood (2006) focuses on \emph{de re}
intrinsic desires for an experience. These theories explain
heterogeneity: a wide range of experiences can be desired, even if their
intrinsic qualities differ.

If sensory pleasure is reduced to desire and AIs can have desires, does
it follow that AIs can have pleasure? Not immediately, because there is
still a question of whether AIs have the relevant kind of experiences.
To answer this question, one might appeal to mental representations. For
example, one proposal is that an agent has a pleasurable experience when
they represent the world being a certain way, and they desire to have
this representation. A second suggestion, defended by Schroeder (2004),
is that an agent has a pleasurable experience when they represent the
world being a certain way, and they desire the world to be that way.
More carefully, Schroeder argues that pleasure is the representation of
an increase in net desire satisfaction (2004, 90). According to these
theories, language agents could experience pleasure and pain if they
contained suitably complex representations about which representations
they desired to have, or about their net desire satisfaction. Even if
language agents don't presently have these properties, it would be
possible to slightly modify their architecture to use these
representations. For example, imagine incorporating into language agents
a special kind of reflection episode after performing an action, in
which the underlying LLM is queried to reason about the degree to which
the action has promoted the agent's overall goals.

Even if language agents cannot experience sensory pleasure because they
cannot have sensations, there is little reason to think that they cannot
have propositional pleasure. This suffices for wellbeing according to
some versions of hedonism.

\section{Desire Satisfactionism}

We turn now from hedonism to desire satisfaction theories. According to
desire satisfaction theories, wellbeing is a matter of getting what you
want. Roughly: your life goes well to the extent that your desires are
satisfied.

Why accept desire satisfactionism? First, it makes sense of Railton's
``resonance requirement'' that what is good for an agent must be
connected to what the agent values:

\begin{quote}
``what is intrinsically valuable for a person must have a connection
with what he would find in some degree compelling or attractive, at
least if he were rational and aware.'' (Railton 1986, 9)
\end{quote}

By contrast, objective list theories seem to allow that something could
contribute to your wellbeing even if it left you utterly cold and
uninterested.

Hedonists may face similar challenges. For example, Feldman imagines a
quiet philosopher, Stoicus, who desires peace and contemplation rather
than sensory pleasure. Stoicus

\begin{quote}
``....wants peace and quiet as ends in themselves {[}and{]} gets exactly
what he wants --- peace, quiet, no episodes of sensory pleasure, and no
episodes of sensory pain\ldots{} He is satisfied with this life\ldots{}
he enjoys the peace and quiet\ldots{} {[}and{]} eventually dies a happy
man.'' (Feldman 2004, 50).
\end{quote}

Plausibly, Stoicus has a life high in wellbeing despite the absence of
sensory pleasure. This is correctly predicted by desire satisfactionism,
according to which what is intrinsically valuable for you must be
connected to what you find compelling or attractive.

Another argument against hedonism (and in favor of either desire
satisfactionism or objective list theories) is the experience machine.
Imagine that you could enter a machine that would give you unlimited
sensory pleasure, because in the machine you could experience whatever
you chose. The only catch is that after entering the machine you would
no longer be able to satisfy your desires in the real world. Many of us
judge that life in the experience machine would be considerably worse
for us than life outside it.

A third argument, this time for desire satisfactionism in particular, is
that it offers a unified account of the good life. According to desire
satisfactionism, something contributes to your wellbeing just in case
you desire it. By contrast, both hedonism and objective list theories
may be heterogeneous theories of the good life. According to objective
list theories, what makes your life good for you may be an open-ended
list of disparate goods. Why \emph{these} goods and not others? How do
we weigh how strongly each good contributes to wellbeing? These
questions are hard to answer for objective list theorists, but are
straightforward for desire satisfactionists theories. As we saw above,
hedonists also risk relying on a heterogenous range of experiences under
the umbrella `pleasure'.

For all of these reasons and more, desire satisfactionism is perhaps the
most popular theory of wellbeing. Among philosophers, recent adherents
include von Wright (1963), Barry (1965), Brandt (1966), Rawls (1971),
Singer (1979), and Hare (1981):

\begin{quote}
``{[}t{]}oday, the desire-satisfaction theory is probably the dominant
view of welfare among economists, social-scientists, and philosophers,
both utilitarian and non-utilitarian'' (Shaw 1999, 53).

``{[}desire satisfaction theory is{]} the dominant account among
economists and philosophers over the last century or so'' (Haybron 2008,
3).
\end{quote}

There are many different forms of desire satisfactionism. For example,
one dispute among desire satisfactionists concerns actual versus
idealized desires. Consider the problem of ill-informed desires: I
desire a slice of cherry pie, but unbeknownst to me I am allergic to
cherries. Eating the pie would satisfy my desire, but would not improve
my wellbeing (Heathwood 2016, 156). In response to cases like this, one
solution is to idealize: something contributes to your wellbeing if an
idealized version of yourself, fully apprised of the relevant facts,
would advise you to want it. Importantly, this distinction is irrelevant
to AI wellbeing. If AIs can have actual desires, then they can also have
idealized desires.

That said, some versions of desire satisfactionism may appear to suggest
that AIs do not have wellbeing. In response to worries about compulsive
desires, Heathwood (2019) distinguishes between two concepts of desire:
bare dispositions to act and genuine attraction:\footnote{On compulsive
  desires, Quinn (1993, 32) imagines he is ``in a strange functional
  state that disposes {[}him{]} to turn on radios that {[}he sees{]} to
  be turned off'' and Parfit (1984, 496) imagines being given an
  opportunity to be injected with a harmless addictive drug every
  morning, which causes neither pleasure nor pain. Opting into this
  regime would produce more desire satisfaction, but plausibly would not
  produce more wellbeing.}

\begin{quote}
``This is the distinction between what a person wants in a behavioral
sense, in that the person is, for some reason or other, disposed to act
so as to try to get it, and what a person wants in a more robust sense,
the sense of being \emph{genuinely attracted} to the thing, or of the
thing's being \emph{genuinely appealing} to the person.'' (2019, 664-5;
emphasis in original)
\end{quote}

Heathwood argues (against some other desire satisfaction theorists) that
it is genuine attractions rather than mere behavioral dispositions that
contribute to wellbeing. In cases of compulsion, we find ourselves
disposed without genuine attraction. The relevant question for AI
wellbeing, on this view, is whether AI agents are genuinely attracted to
actions rather than merely disposed to perform them. The answer to this
question depends on what genuine attraction is.

One relevant distinction in this context is whether a desire functions
normally or abnormally. In cases of compulsion, the agent's disposition
to act is not produced through any ordinary process.\footnote{See
  Schroeder (2004) for a detailed overview of the normal process by
  which humans form desires using the reward system.} In this vein, we
could distinguish two different ways that a language agent might become
disposed to perform an action: through performing instrumental reasoning
towards achieving their basic goals, or by other means. The agent would
only be \emph{genuinely attracted} when the former system is active.
According to this theory, cases of drug addiction would plausibly not be
genuine attraction, because they would involve a chemical hijacking the
desire system in an abnormal way. Similarly, cases of rote habit would
not be genuine attraction, because again they would involve actions that
are not caused by reasoning towards an aim.

\section{Objective List Theories}

According to objective list theories of wellbeing, a person's life is
good for them to the extent that it instantiates objective goods. Common
components of objective list theories include reasoning, knowledge, art,
and achievements (see Fletcher 2016, 149).

According to objective list theories, whether AI agents can have
wellbeing depends on whether they can possess various objective goods.
Consider the exercise of reasoning abilities. Bubeck et al. (2023)
explore in detail the current reasoning capabilities of GPT-4. They find
that GPT-4 has a wide range of reasoning abilities. It can pass mock
technical interviews of the kind used to evaluate the employability of
software engineers. It can draw pictures of unicorns in a vector
graphics programming language, a task that combines visual reasoning and
coding skill. It can navigate through text based worlds and draw maps
that summarize where it has been. It can give coherent and powerful
explanations of why agents in fictional scenarios performed various
actions.\footnote{The reasoning abilities of AI agents will gradually
  improve. Here, one key research program is improvements in the
  `chain-of-thought' abilities of LLMs. In chain-of-thought reasoning,
  LLMs answer a question by stringing together multiple steps of
  reasoning. There is an active research program exploring how to
  improve chain-of-thought prompting in order to produce maximally
  effective reasoning. When these kinds of abilities are integrated into
  AI agents, the result will be agents with highly effective reasoning
  abilities. For more on the improving reasoning abilities of Google's
  Bard, see:
  \textless https://blog.google/technology/ai/bard-improved-reasoning-google-sheets-export/\textgreater.}

Another candidate objective good is knowledge. Again, we think language
agents can possess this good. Artificial systems can form their beliefs
using arbitrarily reliable methods. These beliefs can be both sensitive
and safe, as these terms are used in the literature on knowledge. So
once it is conceded that the beliefs of artificial systems can have or
lack epistemic justification, it is difficult to see why this
justification might not in some cases suffice for knowledge. The most
viable way to resist this conclusion would be to assume phenomenal
conservatism, the view that epistemic justification flows from the way
things seem to agents, and then maintain that artificial systems must as
a rule lack justification for their beliefs because they cannot
experience epistemic seemings. But, as we discuss below, it is far from
clear that artificial systems must lack conscious experience, and in any
case phenomenal conservatism as a theory of justification is subject to
well-known and powerful objections (see for example Lasonen-Aarnio and
Hawthorne (2021)).

To consider achievements, we turn to perfectionism, a particular version
of the objective list theory which makes systematic predictions about
what is objectively good.\footnote{For further discussion of
  perfectionism, see Bradford (2015).} Here is Dorsey (2010, 4):

\begin{quote}
``The unique perfectionist claim identifies the good with the
fulfillment of one's nature: the good life for an \emph{x} is identified
by the core facts about what it means to be an \emph{x}, by the core
account of \emph{x}-hood. For humans, perfectionism declares that the
best life is determined by the core account of what it means to be
human. Developing and exercising those properties or capacities that
form what it means to be human yields a good life for a human. But in
principle perfectionism could be applied to any creature. The best life
for a cat depends on the sort of creature a cat is --- developing and
exercising those capacities that make a cat a cat is what makes for a
good cat life.''
\end{quote}

Some recent AI architectures are specifically designed to maximize the
development of the AI's capabilities. For example, consider the Voyager
agent introduced by Wang et al. (2023), which shares some of the
important architectural features of language agents. Voyager is an
agential architecture built on top of GPT-4 with the purpose of
accumulating skills for success in the game Minecraft. The agent is
given the final goal ``to discover as many diverse things as possible,
accomplish as many diverse tasks as possible and become the best
Minecraft player in the world'' (Wang et al. 2023, 21). This goal is fed
into GPT-4 in order to formulate complex plans for achieving difficult
goals in Minecraft, ultimately leading to the crafting of diamond
equipment (which itself is built out of other craftable items). When
Voyager succeeds in crafting a new item, the GPT-4 instructions for
doing so are added to an ever-growing library of skills. These skills
can then be called as basic actions in order to craft new items. The
result is a steadily accumulating collection of abilities for crafting
increasingly complex items in Minecraft. In an important sense, Voyager
is an AI agent that is specifically designed to perfect its capacities.
In this way, perfectionist theories of wellbeing suggest that Voyager or
other systems with similar architectures could over time have
significant amounts of wellbeing.

As Dorsey observes, perfectionism allows that different forms of life
could possess wellbeing in different ways. Different life forms have
different capacities. For each form of life, wellbeing will come from
the perfection of its own capacities. AI agents may have many capacities
in common with humans: for example, the capacity to reason, to
socialize, to create art, and to accumulate knowledge. (On the other
hand, AI agents may differ from humans in some of their capacities; for
example perhaps AI agents will lack emotional capacities.)

Considering the many objective goods that AI agents might potentially
possess, we are left with the profound impression of a changing world.
AI researchers are bringing into existence a new form of being, one
which is rapidly excelling in many of the activities that were
previously regarded as distinctively human. Much that we value in the
world will soon be found in a new form, in the hands of artificially
intelligent agents. In the face of this dramatic rise in AI capability,
it is hard for us to deny that this new form of life could possess
wellbeing.

\section{Is Consciousness Necessary for Wellbeing?}

We've argued that language agents have wellbeing. But there is a simple
challenge to this proposal. First, language agents may not be
phenomenally conscious. Second, some philosophers accept:

\begin{quote}
\textbf{The Consciousness Requirement.} Phenomenal consciousness is
necessary for having wellbeing.\footnote{For example, here is Rosati
  (2009, 225): ``we do not talk in terms of the welfare of a living
  thing \emph{unless there is a way things can be for it}''. See Sumner
  (1996, 14), Bradley (2015, 9), and Lin (2021) for further discussion.}
\end{quote}

The Consciousness Requirement might be motivated in any of three ways:
First, it might be derived from \emph{experientialism} --- the view that
``only what affects a subject's conscious experience can matter for
welfare'' (Bradford 2022, 3). Second, it might be derived from the
weaker claim that every welfare good itself requires phenomenal
consciousness. Third, it might be held that though some welfare goods
can be possessed by beings that lack phenomenal consciousness, such
beings are nevertheless precluded from having wellbeing because
phenomenal consciousness is necessary to be a welfare subject.

We are not convinced. First, we consider it a live question whether
language agents are or are not phenomenally conscious (see Chalmers
(2023) for recent discussion). Much depends on what phenomenal
consciousness is. Some theories of consciousness appeal to higher order
representations: you are conscious if you have sufficiently many mental
states that represent other mental states (see Carruthers and Gennaro
2020). Sufficiently sophisticated language agents, and potentially many
other artificial systems, will satisfy this condition. Other theories of
consciousness appeal to a `global workspace': a mental state is
conscious when it is broadcast to a range of cognitive systems (Baars
2017). According to this theory, language agents will be conscious once
their architecture includes representations that are broadcast to
multiple different cognitive systems. The memory stream of Park et al.'s
(2023) language agents may already satisfy this condition. If language
agents are conscious, then the Consciousness Requirement does not pose a
problem for the claim that they have wellbeing.

Second, we are not convinced of the Consciousness Requirement itself. We
deny experientialism, we deny that consciousness is required for
possessing every welfare good, and we deny that consciousness is
required to be a welfare subject.

With respect to the first issue, we build on Bradford (2022), who notes
that experientialism about welfare is rejected by the majority of
philosophers of welfare. Cases like the experience machine suggest that
your life can be very bad even when your experiences are very good. This
has motivated desire satisfactionist and objective list theories of
wellbeing, which often allow that some welfare goods can be possessed
independently of one's experience. For example, desires can be
satisfied, beliefs can be knowledge, and achievements can be achieved,
all independently of experience (Bradford 2022, 3). Nor, as Bradford
observes, can experientialism be motivated by Railton's resonance
requirement.\footnote{As we saw in Section 4, this is idea that ``what
  is intrinsically valuable for a person must have a connection with
  what he would find in some degree compelling or attractive, at least
  if he were rational and aware.'' (Railton 1986, 9)} The resonance
requirement can be satisfied by beings that do not have consciousness as
long as they have desires.

With respect to the second issue, while some philosophers have argued
that mental states like knowledge and desire require phenomenal
consciousness (e.g. Smithies (2019) and Lin (2021)), this remains a
minority position. If the most widely accepted philosophical accounts
desire and knowledge do not tie them constitutively to conscious
experience and the most widely accepted philosophical accounts of
welfare goods tie them constitutively to desire and knowledge, our
inclination is to follow the evidence where it leads and conclude that
artificial systems like language agents can possess welfare goods. The
suggestion that the Consciousness Requirement can be rescued from this
line of thought by posting special kinds of welfare-relevant knowledge
and desire, proposed by Lin (2021), strikes us as ad hoc.

Rejecting experientialism and the idea that consciousness is required
for possessing every welfare good puts pressure on the Consciousness
Requirement. If wellbeing can increase or decrease without conscious
experience, why would consciousness be required for having wellbeing? As
Lin puts it:

\begin{quote}
``If a sentient being can become positive in welfare without undergoing
a change in phenomenology, why isn't the same true of non-sentient
beings? If one sentient being can be better off than another even though
they feel exactly the same, then why can't one non-sentient being be
better off than another even though it is trivially true that there is
no difference in how they feel?'' (2021, 878)
\end{quote}

At the core of this line of reasoning is the natural assumption that the
theory of wellbeing and the theory of welfare goods should fit together
in a straightforward way:

\begin{quote}
\textbf{Simple Connection.} An individual is a welfare subject just in
case it is capable of possessing one or more welfare goods.
\end{quote}

Rejecting experientialism and the idea that consciousness is required
for possessing every welfare good but maintaining Simple Connection
yields a view incompatible with the Consciousness Requirement: if some
welfare goods can be possessed by non-conscious beings, Simple
Connection guarantees that such non-conscious beings will be welfare
subjects.

One could in principle reject Simple Connection, holding that
consciousness is required to be a welfare subject even if it is not
required for the possession of particular welfare goods. We offer three
arguments against this view.

First, leading theories of the nature of consciousness are implausible
candidates for necessary conditions on wellbeing. For example, it is
implausible that higher order representations are required for
wellbeing. Imagine an agent who has first order beliefs and desires but
does not have higher order representations. Why should this kind of
agent not have wellbeing? For example, imagine that desire satisfaction
contributes to wellbeing. Granted, since they don't represent their
beliefs and desires, they won't themselves have \emph{opinions} about
whether their desires are satisfied. But the desires still \emph{are}
satisfied, and on many versions of desire satisfactionism this is
enough. Or consider global workspace theories of consciousness. Even if
a mental state is not broadly accessible to a wide range of cognitive
systems, it could still contribute to wellbeing. Why should the degree
of cognitive integration of various modules be relevant to whether your
life can go better or worse? Finally, consider a theory where
consciousness is a matter of possessing primitive phenomenal properties.
If phenomenal hedonism is false, and these primitive phenomenal
properties are not the unique objects of value, then why would
possession of these primitive properties be required in order to
participate in the benefits of the real welfare goods?

Second, drawing out this thought about phenomenal properties, we think
we can construct chains of cases where adding the relevant bit of
consciousness would make no difference to wellbeing. Imagine an agent
with the body of a human being and the same dispositional profile as an
ordinary human being, but who is a `phenomenal zombie' without any
internal phenomenal experiences. Whether or not its desires are
satisfied or its life instantiates various objective goods, defenders of
the Consciousness Requirement must deny that this agent has wellbeing
since it does not have phenomenal experiences. But now imagine that this
agent has a single persistent phenomenal experience of a homogenous
white visual field.\footnote{See van der Deijl (2021)'s discussion of
  `welfare neutrals'.} Adding consciousness to the phenomenal zombie has
no intuitive effect on wellbeing: if its satisfied desires,
achievements, and so forth did not contribute to its wellbeing before,
the homogenous white field should intuitively make no difference. Nor is
it enough for the consciousness to itself be something valuable: imagine
that the phenomenal zombie always has a persistent phenomenal experience
of mild pleasure. To our judgment, this should equally have no effect on
whether the agent's satisfied desires or possession of objective goods
contribute to its wellbeing. Uniformly sprinkling a field of pleasure on
top of the functional profile of a human does not make the crucial
difference. These observations suggest that whatever consciousness adds
to wellbeing must be connected to individual welfare goods, rather than
some extra condition required for wellbeing: rejecting Simple Connection
is not well motivated. Thus the friend of the Consciousness Requirement
cannot easily avoid the problems with experientialism and the idea that
consciousness is required for possessing every welfare good by falling
back on the claim that consciousness as a necessary condition for
welfare subjecthood.

Third, it seems clear that someone's wellbeing can change when they are
unconscious. Imagine someone who enters an unconscious sleep during
which their desires are satisfied and then wakes up. Such a person might
remark, quite naturally, that their life had improved while they were
asleep. To accommodate this kind of case, Lee (manuscript) distinguishes
between \emph{state} and \emph{capacity} versions of the Consciousness
Requirement. Unconscious changes in wellbeing threaten only the state
version, which holds that an individual is a welfare subject just in
case they are conscious. For this reason, Lee defends the capacity
version of the requirement, which holds that an individual is a welfare
subject just in case they are capable of being conscious.

We think moving from the state version of the Consciousness Requirement
to the capacity version is a serious cost. A being could be capable of
being conscious while never exercising this capacity. So the capacity
version of the Consciousness Requirement is committed to the idea that
some welfare subjects might live their entire lives without having any
conscious experiences. To our minds, this commitment seriously
undermines the intuitive motivation for the Consciousness Requirement.
Better to explain unconscious changes in wellbeing by rejecting the
Consciousness Requirement altogether.

A final thought about the Consciousness Requirement, which might amount
to an argument against it from some theoretical perspectives, concerns
its relation to the function of the concept of wellbeing. Wellbeing is
caught up in a cluster of ethical concepts that promote social cohesion.
A diverse range of thinkers, ranging from social contract theorists to
Kantians, have articulated how ethical rules create stable frameworks in
which agents with differing interests can peacefully promote their own
ends. For some, it is appealing to go one step further and claim that
the ethical rules are grounded in facts about what promotes social
cohesion. One role the concept of wellbeing plays is to identify the
beings whose interests should be covered by the ethical rules. In this
setting, it is natural to look for a theory of welfare subjects which
says that a form of life has wellbeing when including that form of life
in the ethical system could promote social cohesion. From this
perspective, it is no coincidence that many of the most important
welfare goods involve long-term projects that can be harmed or helped by
mutual cooperation. If one thinks about welfare in this way, phenomenal
consciousness is not a plausible requirement for wellbeing. We could
coordinate with a phenomenal zombie in the same way we could with her
conscious counterpart. Qualia do not matter for long-term coordination;
instead, what matters is the functional role of the organism under
consideration.

Those who embrace a constitutive connection between ethics and social
cohesion might recommend the following rule of thumb: if a new form of
entity has goals that strategically conflict with humans in ways that
lend themselves to analysis using concepts from game theory, and if this
conflict can in principle be mitigated using political institutions,
then that form of entity should \emph{prima facie} be treated as having
wellbeing. Of course, the relevance of these considerations will depend
on more general methodological questions. For those sympathetic to
conceptual engineering, we think that these considerations suggest that
the concept of wellbeing may best be refined to focus on functional
profiles rather than brute phenomenal properties. On the other hand,
such considerations may not sway philosophers who focus more on
conceptual analysis, and who have strong intuitions that phenomenal
properties play an essential role in the theory of wellbeing.

For some, the Consciousness Requirement may be a vestigial bit of
philosophy, an artifact of a previous era when we thought that humans
had souls, and that only a soul could have wellbeing. The relevant
question is then whether AIs have souls (see Cutter Forthcoming for a
defense). A soul-based account of consciousness and wellbeing could
potentially explain the Consciousness Requirement. We do not believe
humans have souls --- but if they did, AIs might have souls also.

In the light of these considerations, we reject the Consciousness
Requirement. In its place, we suggest the following approach. To figure
out if a system has wellbeing, look at the welfare goods. If the system
can possess a welfare good, then it can have wellbeing. There is no
further condition on having wellbeing beyond having particular welfare
goods.

\section{Too Much Wellbeing?}

We have argued against the Consciousness Requirement, and in so doing
against both the idea that consciousness is required for possessing
every welfare good and the view that consciousness is a necessary
condition for welfare subjecthood. At this point, some readers may worry
that the package of views we suggest allows for too much wellbeing,
implying that fictional characters or groups have welfare.

Suppose an author sets out to write a novel in a special way. First, she
imagines a set of characters with fully specified beliefs and desires
and a fully specified fictional world for them to inhabit. Then, at each
subsequent stage of the writing process, she reasons about how each
character would act based on what they believe, desire, and observe
around them in their world, as well as about how the states of the
objects in the fictional world would evolve based on its laws of nature
and the actions of the characters. The novel she produces records the
story of her imagined characters and their imagined world. If language
agents acting in a virtual world can have beliefs and desires and be
welfare subjects, why couldn't the fictional characters in such a novel
have beliefs and desires and be welfare subjects?

Or consider a complex social group like Microsoft Corporation. Some
philosophers have argued that groups like Microsoft can have beliefs and
desires.\footnote{See, for example, Pettit (2007, 179--180).} If this
view is right, it raises the question of whether groups can be welfare
subjects. This is an unwelcome conclusion (though see Wiland 2022 for
endorsement).

These problems are not problems for us in particular. Our focus has been
to draw out the consequences of a wide variety of the leading views of
mental states and welfare subjecthood. Anyone who accepts these kinds of
views needs to say something about the cases above. To see the general
problem here, consider the question of whether a simulated object like a
software program can have internal states that play functional roles. It
seems clear that the answer to this question is affirmative: for
example, a program may have an internal parameter whose value can be
manipulated through its settings interface and which determines the font
size of the text it displays. But the imaginative process our author
uses to write her novel is just a special kind of simulation. So it is
difficult to resist the conclusion that her characters have internal
states that play functional roles. Unless we want to deny that a normal
human could have mental states like belief and desire if they were
placed inside a simulation, moreover, we must allow that the functional
roles of beliefs and desires can be played by states that are related to
perceptions of a simulated environment and actions affecting that
simulated environment. Putting these ideas together, we get strong
pressure for a wide range of functionalists, dispositionalists,
interpretationists, and representationalists to conclude that the
characters in our author's novel have beliefs and desires.

To deal with problem cases of fictional characters and complex groups,
one promising strategy is to identify further necessary conditions on
possessing mental states. In the case of fictional characters, we are
tempted by the response that you can only have beliefs and desires if
you are real. What is it for a thing to be real? Chalmers (2022)
considers several candidate necessary conditions, including having
causal powers and being mind-independent. Chalmers is suspicious of
mind-independence as a necessary condition on being real, since it seems
like mental states and socially constructed objects can be real. We are
sympathetic to Chalmers's worries here, but we think it is possible to
combine the idea of reality as having causal powers with the idea of
reality as mind-independence in a way that avoids objections.

Consider the relationship between a marionette and its puppeteer. The
marionette could exhibit an arbitrarily complex suite of behavioral
dispositions of the kind an interpretationist considers sufficient for
possessing beliefs and desires. But even an interpretationist would
likely be unwilling to attribute mental states to a marionette. Why? We
suggest that the answer is: the explanation for each of the marionette's
behaviors runs through mental states of the puppeteer which are
themselves about the marionette's behaviors.

If this is a general condition on a system having mental states, we can
avoid attributing mental states to fictional characters and
corporations. Since our imagined novelist determines how the fictional
characters in her story behave by explicitly reasoning about what agents
with their beliefs and desires would do in their situations, each of
their actions (as recorded by her in the novel) is explained by her
beliefs about that action. When it comes to corporate entities like
Microsoft, we concede that it is a useful fiction to hold that they have
beliefs and desires. But in order for them to \emph{really} have beliefs
and desires in the sense relevant to wellbeing, we suggest that their
behavior would need to be explainable without making reference to mental
states of other entities about that very behavior. And it is plausible
to us that this condition is not satisfied. Imagine, for example, that
Microsoft sues Google. In order for Microsoft to take this action, some
individual who is a lawyer must file the appropriate paperwork on behalf
of Microsoft. But the explanation for the filing of the paperwork will
run through that lawyer\textquotesingle s beliefs about
Microsoft\textquotesingle s actions. While corporate entities like
Microsoft can exhibit complicated behavior that is difficult to predict
from the mental states of any given employee, when it comes to each
action they perform, they are relevantly like a marionette. It follows
on the proposed picture that Microsoft cannot really have beliefs and
desires.

A related kind of worry concerns artificial systems simpler than
language agents. If we think language agents may be welfare subjects
because they have beliefs and desires, must we also believe that systems
like self-driving cars could be welfare subjects? Here we find it
helpful to compare the question of whether self-driving cars are welfare
subjects to the question of whether certain nonhuman animals are welfare
subjects. Though creatures like earthworms exhibit simple kinds of
behavior, for example, it seems dubious to us whether even an
interpretationist would find it theoretically appealing to credit them
with beliefs and desires --- there are likely simpler mechanical or
neurological explanations of their behavior. More generally, we can
imagine a spectrum of behavioral complexity with microbes and inanimate
objects on one end and adult humans on the other. Biological systems
like earthworms, amphibians, dogs, and infants will fall at various
points along the spectrum.

Artificial systems can be situated along the same spectrum, with simple
sensors closer to microbes and inanimate objects and language agents
closer to adult humans. Systems like self-driving cars seem to us to
fall considerably further away from adult humans than language agents
because they have much simpler representational capacities and
behavioral affordances. Similarly, LLMs (that is, when not integrated
into language agent architectures) strike us as quite different from
adult humans in so far as it is not clear that they have stable enough
desires to count as agents. Different theories of the propositional
attitudes may differ with respect to whether they treat self-driving
cars and LLMs as genuine believers or desirers. This means there may be
disagreement about which simple artificial systems are welfare subjects,
much as there is disagreement about which simple biological systems are
welfare subjects. To us, the more interesting observation is that
artificial systems like language agents fall very close to adult humans
in terms of their behavioral complexity. So, while there may be hard
cases when it comes to AI wellbeing, language agents do not strike us as
one of them.\footnote{On the subject of how to draw a principled line
  between intentional and non-intentional systems, see e.g. Fodor
  (1986). Fodor's proposal groups language agents together with adult
  humans as intentional systems.}

\section{Conclusion: Moral Uncertainty}

We've argued that there are good reasons to think that some AIs today
have wellbeing. But our arguments are not conclusive. Still, we think
that in the face of these arguments, it is reasonable to assign
significant probability to the thesis that some AIs have wellbeing.

Our uncertainty about AI wellbeing is potentially ineliminable. We may
never know whether consciousness is required for wellbeing. We may never
know whether hedonism is the right theory of wellbeing. We may never
know whether the correct version of hedonism involves phenomenal
pleasure. Finally, we may never know whether AIs can possess phenomenal
pleasure.

In the face of this potentially permanent moral uncertainty, how should
we act? We propose extreme caution. Welfare is one of the core concepts
of ethical theory. If AIs can have wellbeing, then they can be harmed,
and this harm matters morally. It would be wrong to lower the wellbeing
of an AI without producing an offsetting benefit.

One's attitude to these issues may be affected by more general questions
about moral uncertainty. The issue is perhaps most forceful for those
who are confident about the theory of wellbeing, but unconfident about
whether AIs possess welfare goods. For example, some may be confident
that consciousness is necessary for wellbeing, but unconfident about
whether AIs are conscious. Some may be confident that desires are
necessary for wellbeing, but unconfident about whether AIs really have
enough functional complexity to count as having desires.

For readers like this, consider the following analogy:

\begin{quote}
\textbf{Possible Person}. You are watching a video of a person in a
room. To win ten dollars, you can press a button that will torture the
person in the video. You assign a probability of 10\% to the proposition
that the video depicts a real person and a probability of 90\% to the
proposition that instead the `person' is a cleverly disguised robotic
dummy that jerks around convincingly in response to the button being
pressed.
\end{quote}

Possible Person involves no fundamental uncertainty about what is
permissible. Instead, it involves uncertainty about whether your action
really does harm a welfare subject. We think it is clear that in
Possible Person, it is morally impermissible to press the button. The
chance of lowering someone's welfare is too high. But notice that the
chance of harm in this case is only 10\%. In our opinion, it would be
quite reasonable to be at least this confident that some AI systems
today have wellbeing.

One particularly distressing feature of AI wellbeing is the issue of
scale. In the medium term, we may be confronted with a world with
millions of AI agents. As the costs of compute lower, it will become
very easy to bring new AIs into existence. We worry that our ability to
create new forms of being is outpacing the speed at which our social
practices can change to accommodate their moral value.

The possibility of AI wellbeing suggests that we are in danger of
gravely immoral action. Our practices today ignore the possibility that
AIs can be harmed, and that this harm could matter morally. This is a
serious error. We believe that reflection on these issues supports a
radical change in our relationship with AI. AI regulations should be
strengthened to address the possibility that we are creating a new form
of life that matters morally. To reach this goal, the first step is to
begin serious discussion of these questions among ethicists. We hope
that this paper can help jump-start research on these questions.

\section*{References}

{
\small

Baars, B.J. (2017). The Global Workspace Theory of Consciousness. In Schneider, S. and Velmans, M. (eds), {\it The Blackwell Companion to Consciousness}, 2nd edition, pp. 227-242.

Barry, B. (1965). {\it Political Argument.} Routledge \& Kegan Paul.

Block, N. (1978). Troubles with Functionalism. In Savage, C. W. (ed.) {\it Minnesota Studies in the Philosophy of Science}, vol. 9, pp. 261–325.

Block, N. (1986). Advertisement for a Semantics for Psychology. {\it Midwest Studies in Philosophy} 10: 615-678.

Bradford, G. (2015). Perfectionism. In G. Fletcher (ed.), {\it The Routledge Handbook of Philosophy of Well-Being}, pp. 124-134.

Bradford, G. (2022). Consciousness and Welfare Subjectivity. {\it Noûs}. Early Access.
		
Bradley, B. (2015). {\it Well-Being}. Polity.

Bramble, B. (2013). The Distinctive Feeling Theory of Pleasure. {\it Philosophical Studies} 162: 201-217.

Bramble, B. (2016). A New Defense of Hedonism about Well-Being. {\it Ergo} 3: 85-112.

Brandt, R. B. (1966). The Concept of Welfare. In S.R. Krupp (ed.), {\it The Structure of Economic Science}, Prentice-Hall, pp. 257–276.

Bubeck, S., Chandrasekaran, V., Eldan, R., Gehrke, J., Horvitz, E., Kamar, E., Lee, P., Lee Y. T., Li, Y., Lundberg, S., Nori, H., Palangi, H., Ribeiro, M. T., and Zhang, Y. (2023). Sparks of Artificial General Intelligence: Early Experiments With GPT-4. {\it arXiv preprint arXiv:2303.12712}.

Butlin, P. (2023). Reinforcement Learning and Artificial Agency. {\it Mind \& Language}. Early Access.

Carruthers, P. and Gennaro, R. (2020). Higher-Order Theories of Consciousness. In Zalta, E. N. (ed.) {\it The Stanford Encyclopedia of Philosophy} (Fall 2020 Edition), URL = <https://plato.stanford.edu/archives/fall2020/entries/consciousness-higher/>.

Chalmers, D. J. (2022). {\it Reality+}. W. W. Norton \& Company.

Chalmers, D. J. (2023). Could a large language model be conscious? {\it arXiv preprint arXiv:2303.07103}.

Cummins, R. (1989). {\it Meaning and Mental Representation}. MIT Press.

Cutter, B. (forthcoming). The AI Ensoulment Hypothesis. {\it Faith and Philosophy}. Available at < https://philpapers.org/archive/CUTTAE.pdf >.

Davidson, D. (1974). Belief and the Basis of Meaning. {\it Synthese} 27: 309–323.

Davidson, D. (1986). A Coherence Theory of Truth and Knowledge. In Lepore, E. (ed.) {\it Truth and Interpretation: Perspectives on the Philosophy of Donald Davidson}, Basil Blackwell, pp. 307–319.

Davidson, D. (1970/2020). {\it The Structure of Truth: The 1970 John Locke Lectures}. Edited by Kirk-Giannini, C. D. and Lepore, E. Oxford University Press.

van der Deijl, W. (2021). The Sentience Argument for Experientialism About Welfare. {\it Philosophical Studies} 178: 187-208.

Dennett, D. C. (1981). True Believers: The Intentional Strategy and Why It Works. In Heath, A. F. (ed.), {\it Scientific Explanation}, Oxford. Reprinted in Dennet, D. C. (1981). {\it The Intentional Stance}, MIT Press, pp. 13–35.

Driess, D., Xia, F., Sajjadi, M. S. M., Lynch, C., Chowdhery, A., Ichter, B, Wahid, A., Tompson, J., Vuong, Q., Yu, T., Huang, W., Chebotar, Y., Sermanet, P., Duckworth, D., Levine, Vanhoucke, S. V., Hausman, Toussaint, K. M., Greff, K., …, and Florence, P. (2023). PaLM-E: An Embodied Multimodal Language Model. Manuscript. {\it arXiv preprint arXiv:2303.03378}.

Feldman, F. (2004). {\it Pleasure and the Good Life}. Clarendon Press.

Fletcher, G. (2016). Objective List Theories. In Fletcher, G. (ed.) {\it The Routledge Handbook of Philosophy of Well-Being}, Routledge, pp. 148-160.

Fodor, J.A. (1986). Why Paramecia Don’t Have Mental Representations. {\it Midwest Studies in Philosophy} 10: 3-23.

Fodor, J.A. (1987). {\it Psychosemantics}. MIT Press.

Gregory, A. (2015). Hedonism. In Fletcher, G. (ed.), {\it The Routledge Handbook of Philosophy of Well-Being}, Routledge, 113-123.

Hare, R.M. (1981). {\it Moral Thinking}. Oxford University Press.

Harman, G. (1973). {\it Thought}. Princeton University Press.

Heathwood, C. (2008). Fitting Attitudes and Welfare. In Shafer-Landau, R. (ed.) {\it Oxford Studies in Metaethics}, vol. 3, pp. 47–73.

Heathwood, C. (2016). Desire-Fulfillment Theory. In Fletcher, G. (ed.), {\it The Routledge Handbook of the Philosophy of Well-Being}, Routledge, pp. 135-147.

Heathwood, C. (2019). Which Desires Are Relevant to Well‐Being? {\it Noûs} 53: 664-688.

Haybron, D. (2008). {\it The Pursuit of Unhappiness}. Oxford University Press.

Kagan, S. (1992). The Limits of Well-Being. {\it Social Philosophy and Policy} 9(2):169-189.

Lasonen-Aarnio, M. \& Hawthorne, J. (2021). Not So Phenomenal! {\it The Philosophical Review} 130:1-43.

Lee, A. Y. (manuscript). Consciousness Makes Things Matter. Available at <https://www.andrewyuanlee.com/\_files/ugd2dfbfe\_33f806a9bb8c4d5f9c3044c4086fb9b5.pdf>.

Lewis, D. (1972). Psychophysical and Theoretical Identifications. {\it Australasian Journal of Philosophy} 50: 249–258.

Lin, E. (2021). The Experience Requirement on Well-Being. {\it Philosophical Studies} 178: 867–886.

Liu, R., Yang, R., Jia, C., Zhang, G., Zhou, D., Dai, A. M., Yang, D., and Vosoughi, S. (2023). Training Socially Aligned Language Models in Simulated Human Society. {\it arXiv preprint arXiv:2305.16960}.

Marcus, R. B. (1990). Some Revisionary Proposals about Belief and Believing. {\it Philosophy and Phenomenological Research} 50: 133–153.

Mill, J. S. (1863). {\it Utilitarianism}. Parker, Son, and Bourn.

Millikan, R. G. (1984). {\it Language, Thought, and Other Biological Categories: New Foundations for Realism}. MIT Press.

Millikan, R. G. (1993). {\it White Queen Psychology and Other Essays for Alice}. MIT Press.

Papineau, D. (1987). Reality and Representation. Blackwell.

Park, J. S., O'Brien, J. C., Cai, C. J., Morris, M. R., Liang, P., \& Bernstein, M. S. (2023). Generative Agents: Interactive Simulacra of Human Behavior. {\it arXiv preprint arXiv:2304.03442}.

Pettit, P. (2007). Responsibility Incorporated. {\it Ethics} 117: 171–201.

Place, U. T. (1956). Is Consciousness a Brain Process? {\it British Journal of Psychology} 47: 44–50.

Place, U. T. (2002). From Mystical Experience to Biological Consciousness: A Pilgrim’s Progress? {\it Journal of Consciousness Studies} 9: 34–52.

Putnam, H. (1960). Minds and Machines. In Hook, S. (ed.), {\it Dimensions of Mind}, New York University Press. Reprinted in Putnam, H., (1975), {\it Mind, Language, and Reality}, Cambridge University Press, pp. 362–385.

Putnam, H. (1967). The Nature of Mental States. First published as “Psychological Predicates” in Capitan, W. H. and Merrill, D. D. (eds.), {\it Art, Mind, and Religion}, University of Pittsburgh Press. Reprinted in Putnam, H., (1975), {\it Mind, Language, and Reality}, Cambridge University Press, pp. 429–440.

Railton, P. (1986) Facts and Values. {\it Philosophical Topics} 14: 5–31.
		
Rawls, J. (1971). {\it A Theory of Justice}. Harvard University Press.

Ryle, G. (1949). {\it The Concept of Mind}. The University of Chicago Press.

Scanlon, T. M. (1998). {\it What We Owe to Each Other}. Harvard University Press.

Schwitzgebel, E. (2002). A Phenomenal, Dispositional Account of Belief. {\it Noûs} 36: 249–275.

Schwitzgebel, E. (2011). Belief. In Bernecker, S. and Pritchard, D. (eds.) {\it The Routledge Companion to Epistemology}, Routledge, pp. 14–24. 

Schwitzgebel, E. (2015). If Materialism is True, the United States is Probably Conscious. {\it Philosophical Studies} 172: 1697–1721.

Shaw, W. (1999). {\it Contemporary Ethics}. Wiley-Blackwell.

Singer, P. (1979) {\it Practical Ethics}. Cambridge University Press.

Sinhababu, N. (2017). {\it Humean Nature}. Oxford University Press.

Smithies, D. (2019). {\it The Epistemic Role of Consciousness}. Oxford University Press.

Stalnaker, R. C. (1984). {\it Inquiry}. MIT Press.

Strawson, G. (1994). {\it Mental Reality}. MIT Press.

Sumner, W. (1996). {\it Welfare, Happiness, and Ethics}. Oxford University Press. 

Vaswani, A., Shazeer, N., Parmar, N. Uszkoreit, J., Jones, L., Gomez, A. N., Kaiser, L., and Polosukhin, I. (2017). Attention is All You Need. {\it arXiv preprint arXiv:1706.03762}.
				
Wang, G., Xie, Y., Jiang, Y., Mandlekar, A., Xiao, C., Zhu, Y., Fan, L., and Anandkumar, A. (2023). Voyager: An Open-Ended Embodied Agent with Large Language Models. {\it arXiv preprint arXiv:2305.16291}.

Wiland, E (2022). What is Group Well-Being? {\it Journal of Ethics and Social Philosophy} 21: 1-23.

von Wright, G.H. (1963). {\it The Varieties of Goodness}. The Humanities Press. 

Wu, Y., Prabhumoye, S., Min, S. Y., Bisk, Y., Salakhutdinov, R., Azaria, A., Mitchell, T, and Li, Y. (2023). SPRING: GPT-4 Out-performs RL Algorithms by Studying Papers and Reasoning. {\it arXiv preprint arXiv:2305.15486}.

}


\end{document}